\documentclass{moriond}

\def\Journal#1#2#3#4{{#1} {\bf #2}, #3 (#4).}

\def\NAT{\em Nature}
\def\JHEP{\em JHEP}

\def\NIMA{{\em Nucl. Instrum. Methods} A}

\def\PLB{{\em Phys. Lett.}  B}
\def\PRL{\em Phys. Rev. Lett.}
\def\PRD{{\em Phys. Rev.} D}
\def\PRP{\em Phys. Rep.}

\def\EPJ{{\em Eur. Phys. J.} C}
\def\CPC{\em Comput. Phys. Commun.}

\def\be{\begin{equation}}
\def\ee{\end{equation}}
\def\bea{\begin{eqnarray}}
\def\eea{\end{eqnarray}}

\usepackage{relsize}
\def\babar{\mbox{\slshape B\kern-0.1em{\smaller A}\kern-0.1em
    B\kern-0.1em{\smaller A\kern-0.2em R}}}

\def\pho{\textsc{Phokhara}}
\def\afk{\textsc{AfkQed}}
\def\GeVM{~${\rm GeV}/c^2$}

\def\GeVP{~${\rm GeV}/c$}
\def\fb{~${\rm fb}^{-1}$}

\newcommand{\mumug}{$\mu\mu\gamma$}
\newcommand{\pipig}{$\pi\pi\gamma$}
\newcommand{\KKg}{$KK\gamma$}
\newcommand{\eeg}{$ee\gamma$}

\newcommand{\eepipig}{$e^+e^- \rightarrow \pi^+\pi^-(\gamma)$}
\newcommand{\eemumug}{$e^+e^- \rightarrow \mu^+\mu^-(\gamma)$}

\usepackage{subfig}

\newcommand{\Photo}{}

\begin{document}
\vspace*{4cm}
\title{Measurement of hadronic cross sections via initial state radiation at \babar}

\author{ Léonard Polat, on behalf of the \babar\ collaboration }

\address{LPNHE, 4 place Jussieu, 75252 Paris Cedex 05, France\\
IJCLab, 15 rue Georges Clémenceau, 91405 Orsay, France}

\maketitle\abstracts{
The \babar\ experiment participates to the global endeavor for a precise prediction of the anomalous magnetic moment of the muon by evaluating the contribution of hadronic processes to the vacuum polarization. After its last result published in 2009 and 2012, \babar\ is preparing a new independent measurement of the \eepipig\ cross section via initial state radiation, with full data statistics and improved uncertainties. A first milestone was reached with the recent study of additional radiations in \eepipig\ and \eemumug, which uncovered shortcomings of the \pho\ Monte Carlo generator in one-photon rates and angular distributions. This has practically no effect on the previous \babar\ measurement, but could explain longstanding discrepancies with other experiments.
}

\section{Introduction}

The anomalous magnetic moment of the muon $a_\mu$ is sensitive to hadronic vacuum polarization (HVP), which is the dominant source of uncertainty on its predicted value~\cite{ta}. It is therefore of great interest for physicists to improve the precision on the HVP contribution to $a_\mu$, in the search for potential tensions with direct measurements.

This contribution can be obtained through a dispersion integral by measuring the cross sections of $e^+e^- \rightarrow \mathrm{hadrons}$ processes. The largest input comes from $e^+e^- \rightarrow \pi^+\pi^-$ and has been measured by many experiments using the initial state radiation (ISR) method (BESIII~\cite{bes}, KLOE~\cite{kloe}, \babar~\cite{babar1,babar2}, CLEO-c~\cite{cleo}) or direct energy scans (SND~\cite{snd}, CMD-3~\cite{cmd})~\footnote{Only some of the most recent results are quoted.}.

\begin{figure}[ht]
\centering
\subfloat[]{\includegraphics[width=0.45\columnwidth]{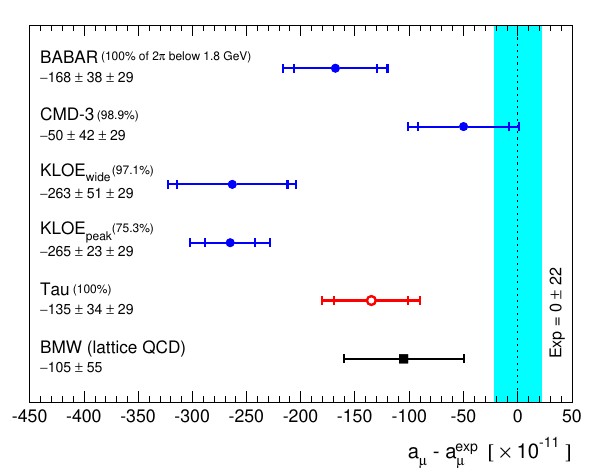}}
\hspace{0.4cm}
\subfloat[]{\includegraphics[width=0.45\columnwidth]{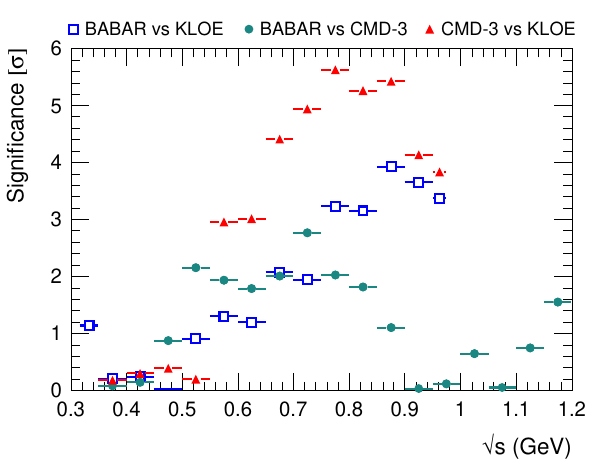}}
\vspace{-0.2cm}
\caption{
Compilation of $a_\mu$ predictions subtracted by the central value of the experimental world average (a) and significance of the difference between \babar, KLOE and CMD-3 for narrow energy intervals of 50 MeV or less (b).~\protect\cite{dhlmz}}
\label{fig:amu}
\end{figure}

As of today, the prediction from the dispersion approach is in tension with the direct measurements of $a_\mu$ (up to a significance of $5\sigma$~\cite{ag}) and with the calculation from lattice QCD (around $2\sigma$~\cite{sb}). Tensions exist as well between different experiments that measured the dipion cross section, especially KLOE and CMD-3, up to more than $5\sigma$ as illustrated in Figure~\ref{fig:amu}. Thus, it is necessary to conduct further measurements to solve these discrepancies.

The last \babar\ result was published in 2009 and 2012. An upcoming analysis, involving a new cross section measurement method, intends to improve its precision using the full data samples collected by the experiment.

\section{The \texorpdfstring{\babar}{BaBar} data and Monte Carlo simulation generators}
\label{sec:babar}

\babar\ is an experiment that operated from 1999 to 2008 at the SLAC National Accelerator Laboratory (USA). It exploited asymmetric collisions of electrons and positrons injected in the storage rings of the PEP-II facility, with respective energies of 9 and 3 GeV and total center-of-mass (c.m.) energy $\sqrt{s}=10.58$\GeVM, at the $\Upsilon(4S)$ meson resonance~\cite{exp}. The experiment collected 424.2\fb\ of data at this resonance and 43.9\fb\ off-resonance~\cite{lumi}.

Collision events are reconstructed with the information provided by the \babar\ detector. The component closest to the beam-beam interaction point is a silicon vertex tracker, followed by a central drift chamber (DCH), both making up the charged particle tracking system. Outside of the DCH are a detector of internally reflected Cherenkov light and an electromagnetic calorimeter (EMC), all within a superconducting solenoid producing a 1.5 T magnetic field. Finally, an instrumented flux return serves to identify muons and detect neutral hadrons.

In addition to the collected data, Monte Carlo (MC) simulation samples are generated for signal events, namely the $\pi^+\pi^-\gamma_{\mathrm{ISR}}$ and $\mu^+\mu^-\gamma_{\mathrm{ISR}}$ final states, where $\gamma_{\mathrm{ISR}}$ stands for the leading order (LO) ISR photon. The employed MC generators are
\begin{itemize}
    \item \pho~\cite{pho}, used to produce signal samples equivalent to 10 times the total \babar\ data luminosity, with full next-to-leading order (NLO) simulation at ISR level: additional ISR photons can be emitted at small or large angles from the beams and interference between ISR and final state radiations (FSR) are taken into account;
    \item \afk~\cite{afk}, used to generate smaller samples (around half of the data luminosity), with NLO and next-to-next-to-leading order (NNLO) simulation at ISR level but neither large angle ISR, nor ISR-FSR interference at NLO and higher orders.
\end{itemize}

Background samples that simulate the processes $e^+e^- \rightarrow q\bar{q}~(q=u,d,s,c),~\tau^+\tau^-,~X\gamma_{\mathrm{ISR}} \\ (X=K^+K^-,~n\pi/K+m\pi^0,...)$ are also generated with \pho, \afk, JETSET~\cite{jet} and KK2f~\cite{kk2f}.

\section{Past and future \texorpdfstring{\eepipig}{e+e- -> pi+pi-(gamma)} cross section measurements at \texorpdfstring{\babar}{BaBar}}

The last \babar\ analysis measured the \eepipig\ cross section as a function of the reduced energy $\sqrt{s'}=m_{\pi^+\pi^-(\gamma)}$, including additional FSR photons, using 232\fb\ of data at the $\Upsilon(4S)$ resonance. The dispersion relation that formulates the lowest-order loop contribution of the $\pi^+\pi^-(\gamma)$ intermediate state to $a_\mu$ is
\begin{equation}
\label{eq:int_amu}
    a_\mu^{\pi\pi(\gamma),\mathrm{LO}} ~=~ 
        \frac{1}{4\pi^3}
        \int\limits_{4m_\pi^2}^\infty ds'\,K(s')\,\sigma^{0}_{\pi\pi(\gamma)}(s')~,
\end{equation}
where $\sigma^{0}_{\pi\pi(\gamma)}(s')$ is the bare cross section (excluding vacuum polarization) of the \eepipig\ process and $K(s')$ is a QED kernel.

To cancel out the ISR photon efficiency and the vacuum polarization, the measured $\pi^+\pi^-(\gamma)$ mass spectrum is divided by the $\mu^+\mu^-(\gamma)$ spectrum, equivalent to the ratio of each final state's bare cross section. The separation between charged pion and muon tracks was based on particle identification (PID), which required the selection $p>1$\GeVP\ on each track momentum to make the muon identification more reliable.

The final prediction of $a_\mu$ obtained by \babar,
\begin{equation}
 a_\mu^{\pi\pi(\gamma),\mathrm{LO}} ~=~ (514.09\pm2.22\pm3.11)\times 10^{-10}~,
\end{equation}
provides first a statistical error, followed by the total systematic uncertainty. The latter covers multiple sources of systematic biases, the PID among the dominant ones. The total relative systematic uncertainty for energies between 0.5 and 1 GeV, that is a large window around the $\rho$ meson resonance peak, is 0.5\%.

The upcoming \babar\ analysis, intended to be published by the end of 2024 or early 2025, aims at improving both the statistical and systematic precision on the $a_\mu$ prediction. For that purpose, all the data collected by the experiment at the $\Upsilon(4S)$ energy (on- and off-resonance) will be studied, while PID requirements on the tracks will be removed.

An angular fit is considered as a new method to distinguish the main signal and background processes (\pipig, \mumug, \KKg, \eeg), based on the cosine of the angle between the negative charge track and the ISR photon in the $2\pi$ c.m.\ frame, assuming both tracks have pion masses. For this purpose, the $p>1$\GeVP\ selection on track momenta required for $\pi/\mu$ identification in the 2009 analysis is released, increasing the statistics at the same time.

Overall, the new analysis will result in an independent measurement of the \eepipig\ cross section, allowing to check the last $a_\mu$ prediction with improved precision. A first step towards this objective was taken with a study of additional radiation in ISR processes, published in 2023~\cite{isr}.

\section{Measurement of additional radiation in initial-state-radiation processes}

The intent of this study is to measure the relative proportions of additional radiations at LO, NLO and NNLO in \eepipig\ and \eemumug\ processes with ISR, as well as to evaluate how accurate the \pho\ and \afk\ generators are in describing the data.

The analysis relies on 424.2\fb\ of data collected on the $\Upsilon(4S)$ resonance energy and 43.9\fb\ off-resonance. Signal simulation samples from \pho\ and \afk\ generators and background samples are described in Section~\ref{sec:babar}. To be consistent with the upcoming \eepipig\ cross section measurement analysis, the tracks are assumed to have pion masses. Dipion and dimuon events are identified according to tight PID selections.

\subsection{NLO fits}

To characterize the possible configurations with a single additional photon, events go through two different fits, depending on whether
\begin{itemize}
    \item it is emitted at large angle from the beams (\textit{LA fit}): the photon must be detected in the EMC with a polar angle between 0.35 and 2.4 rad and an energy larger than 50 MeV, both quantities being used in the fit,
    \item or it is emitted at small angle from the beams (\textit{SA fit}): no information is measured, therefore the photon is simply assumed to be collinear with one of the beams while any other photon detected in the EMC is ignored.
\end{itemize}
Other inputs to the fits are the measured energy and direction of the ISR photon, as well as the momenta and angles of both charged tracks. The fits return $\chi^2$ values that quantify their goodness, along with fitted kinematic variables. Events are classified as NLO LA or NLO SA depending on the fit that provides the smallest $\chi^2$, with the additional requirement that fitted energies are larger than 200 MeV either in the laboratory frame for NLO LA events ($E_{\gamma_{\mathrm{LA}}}$) or in the $e^+e^-$ c.m.\ frame for NLO SA events ($E^*_{\gamma_{\mathrm{SA}}}$). Events that don't pass these energy thresholds are classified as LO.

Significant amounts of background events contaminate the samples, especially in the \eepipig\ process. They are rejected with selection criteria defined on the two-dimensional NLO SA $\chi^2$ vs NLO LA $\chi^2$ plane, optimized with boosted decision trees in three large mass regions. These 2D selections retain between 98 and 99\% of signal depending on the channel and the mass window.

\subsubsection{Additional photon at large angle (NLO LA)}

FSR photons are typically emitted at large angles from the beams, which is also the case of a non-negligible portion of ISR photons. The distinction is performed by fitting the distribution of the minimum angle $\theta_{\mathrm{min(trk,\gamma_{LA})}}$ between the additional $\gamma_{\mathrm{LA}}$ photon and the charged tracks.

MC samples generated by \afk\ have no LA ISR photons and are therefore used to get a template for FSR photons. The template for LA ISR photons is then derived from \pho\ generated samples, subtracting the FSR template given by \afk. By fitting the data $\theta_{\mathrm{min(trk,\gamma_{LA})}}$ distribution, the separation between FSR and LA ISR is fixed at 20 degrees in both \pipig\ and \mumug\ samples. Indeed, the angle between FSR photons and closest tracks peaks between 5 and 10 degrees, followed by a steep decrease, while the LA ISR component forms a wide bump centered around 60 degrees.

\afk\ generated samples are observed to be in good agreement with data as a function of the $\gamma_{\mathrm{LA}}$ energy. The same conclusion applies to \pho, below and above 20 degrees.

\subsubsection{Additional photon at small angle (NLO SA)}

\begin{figure}[ht]
\centering
\subfloat[]{\includegraphics[width=0.45\columnwidth]{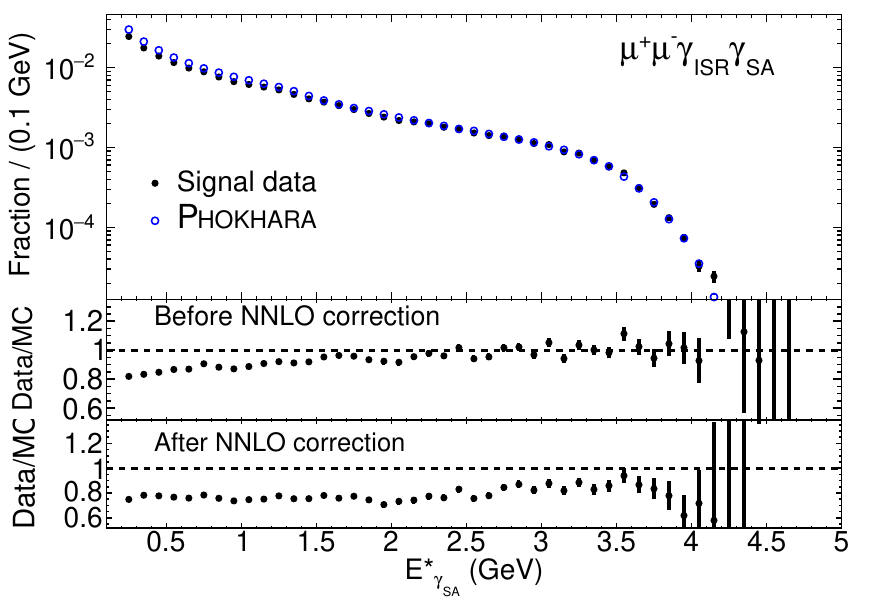}}
\hspace{0.4cm}
\subfloat[]{\includegraphics[width=0.45\columnwidth]{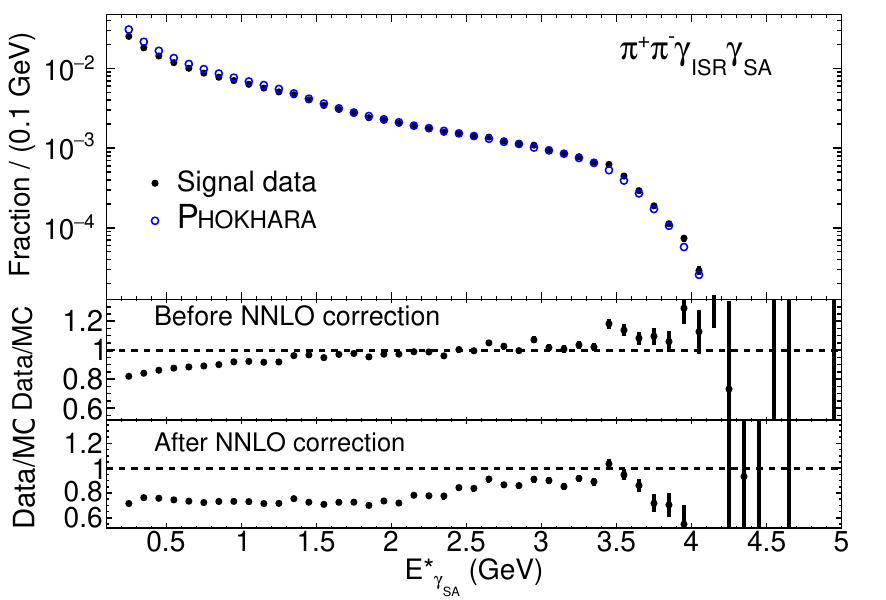}}
\vspace{-0.2cm}
\caption{
Comparison of the fitted energy $E^*_{\gamma_{\mathrm{SA}}}$ in the c.m.\ frame of the additional NLO SA photon, before and after NNLO correction, in the \mumug\ (a) and \pipig\ (b) data and \pho\ samples.}
\label{fig:NLO_SA} 
\end{figure}

Virtually all radiations emitted close to the beams correspond to ISR photons. In contrast with the NLO LA fits, an excess of NLO SA events is observed in \pho\ samples compared to data, evolving with a positive slope as a function of the additional $\gamma_{\mathrm{SA}}$ photon's energy in the c.m.\ frame, as shown in Figure~\ref{fig:NLO_SA}.

\begin{figure}[ht]
\centering
\subfloat[]{\includegraphics[width=0.45\columnwidth]{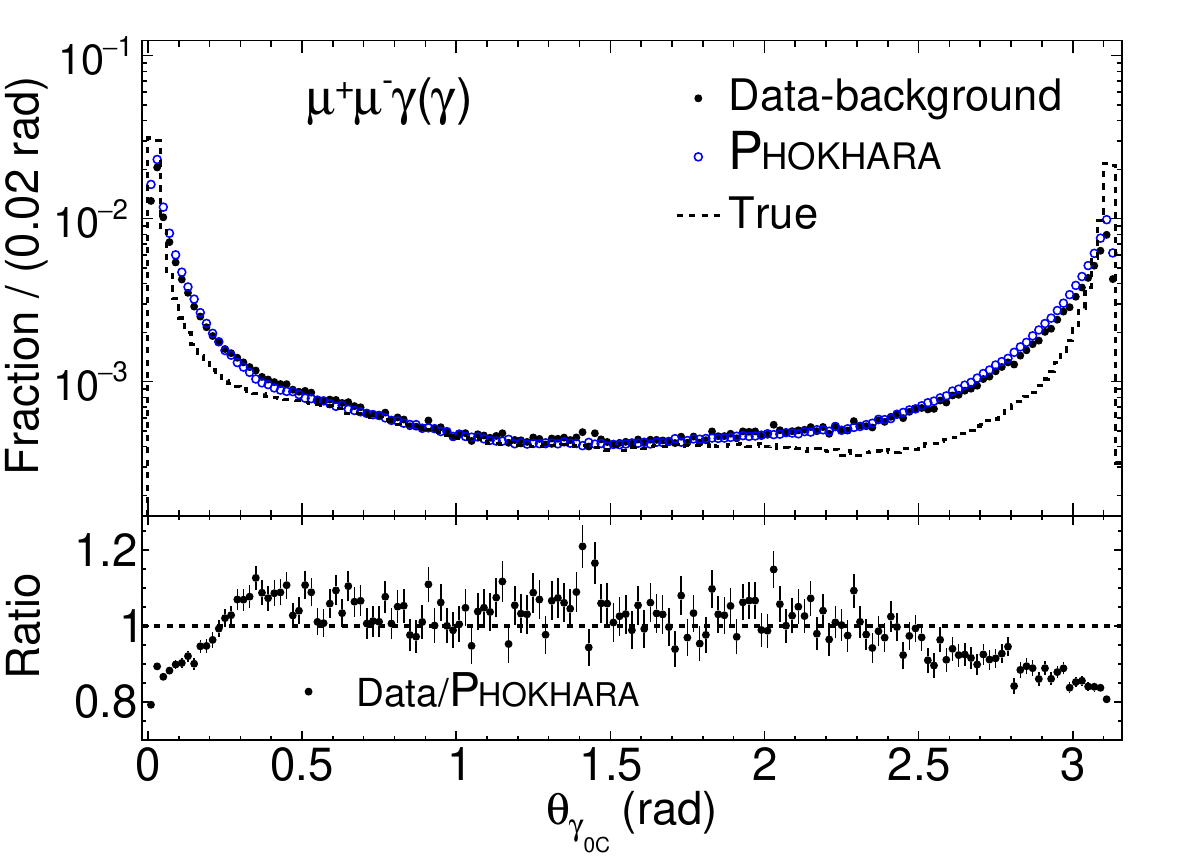}}
\hspace{0.4cm}
\subfloat[]{\includegraphics[width=0.45\columnwidth]{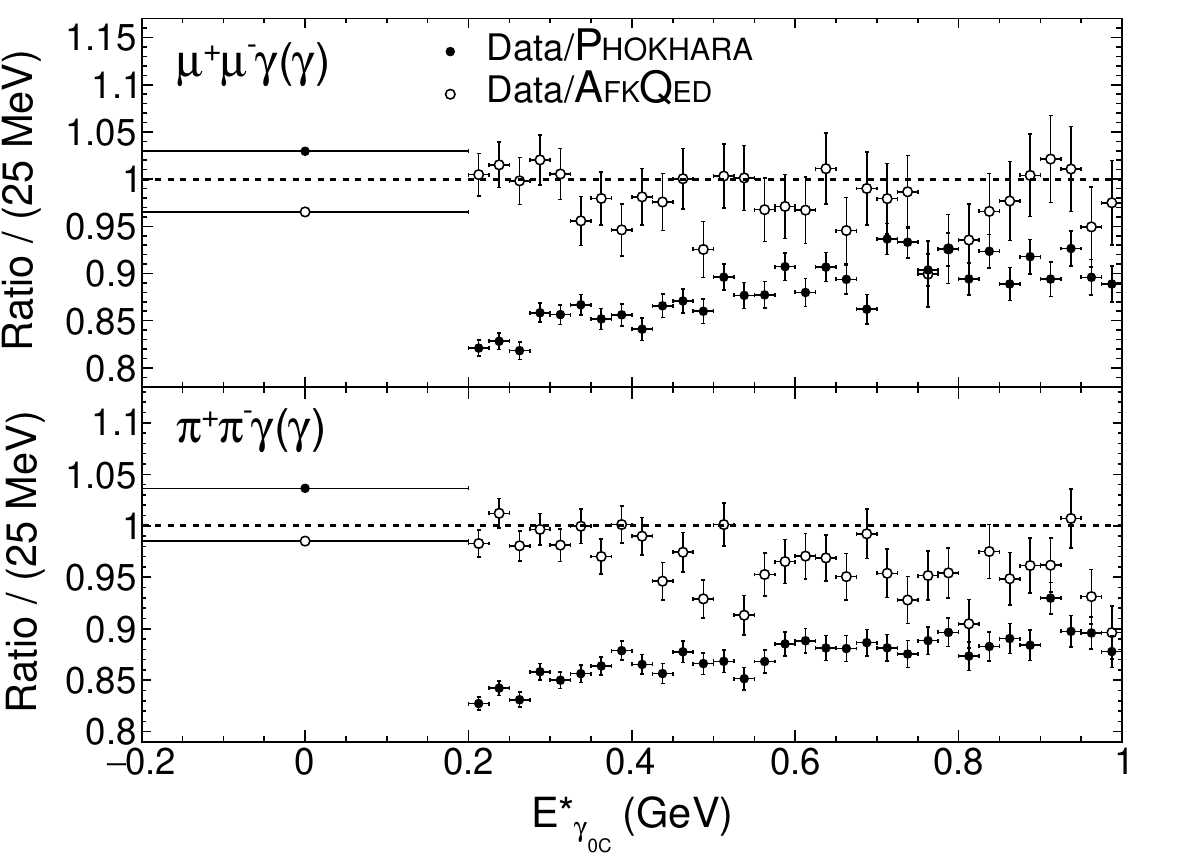}}
\vspace{-0.2cm}
\caption{
(a) Distributions of $\theta_{\gamma_\mathrm{0C}}$ for dimuon events with $E^\ast_{\gamma_\mathrm{0C}}>0.2$ GeV in data and \pho, along with MC truth information. (b) Data/\pho\ and data/\afk\ ratios of $E^\ast_{\gamma_\mathrm{0C}}$ distributions, in dimuon (top) and dipion (bottom) samples.}
\label{fig:NLO_0C} 
\end{figure}

A possible reason could be the assumption that small-angle photons are exactly collinear to the beams, as required in the NLO SA fits. To test this hypothesis, an alternative zero-constraint (0C) calculation of the angle $\theta_{\gamma_\mathrm{0C}}$ and energy in the c.m.\ frame $E^\ast_{\gamma_\mathrm{0C}}$ of the additional photon is performed, without assuming any collinearity. Figure~\ref{fig:NLO_0C} shows that events with a small-angle additional photon are overestimated by the \pho\ simulation. The trend observed in the $\gamma_{\mathrm{SA}}$ energy distribution remains visible in the $\gamma_{\mathrm{0C}}$ photon energy, with a complete mismatch between LO ($E^*_{\gamma_{\mathrm{0C}}}<200$ MeV) and NLO ($E^*_{\gamma_{\mathrm{0C}}}>200$ MeV) events. This is not the case in \afk, which agrees with data along the full energy range.

Therefore, the 0C calculation shows that the excess of \pho\ events with small-angle additional photons is not due to the collinear assumption in the NLO SA fits. It is, however, affected by the presence of NNLO photons as demonstrated in Section~\ref{subsec:corr_SA}.

\subsection{NNLO fits}

NNLO fits are performed only on data and \afk\ generated samples, since \pho\ doesn't simulate more than a single additional photon. Like the NLO fits, they categorize events depending on whether they have additional photons emitted at large or small angles from the beams, in this case: two small-angle (\textit{2SA}), two large-angle (\textit{2LA}) or one small-angle and one large-angle (\textit{SA+LA}) photons. Events are assigned to the category that provides the smallest $\chi^2$ compared to all the others, including NLO fits.

Significant NNLO contributions are found after subtracting the background from NLO events, evaluated on \pho\ samples. The dominant configuration corresponds to two additional ISR photons (2SA or SA+LA), amounting to more than 3\% of LO, NLO and NNLO categories combined. The total proportions of NNLO events are $(3.36 \pm 0.39)\%$ and $(3.47 \pm 0.38)\%$ in the pion and muon channels, respectively.

\afk\ shows overall good performance in describing the rates and energy distributions of data at NNLO. A slightly high data/MC ratio of $1.061 \pm 0.015$ for muons and $1.043 \pm 0.010$ for pions is found up to the maximum generated energy.

\subsection{NNLO correction to NLO SA results}
\label{subsec:corr_SA}

The NLO SA category in data is impacted by NNLO 2SA feed-through: two small-angle NNLO photons emitted from the same beam cannot be distinguished from a single NLO SA photon. Figure~\ref{fig:NLO_SA} shows that there is a better agreement in shape with \pho\ after the $E^*_{\gamma_{\mathrm{SA}}}$ distributions in data are corrected to take this effect into account. However, an almost constant excess of simulated events is observed, with an overall data/MC ratio of $0.763 \pm 0.019$ in \eepipig\ and $0.750 \pm 0.008$ in \eemumug.

\section{Conclusion}

The study of additional radiation in ISR processes shows a significant contribution of NNLO radiations in \eepipig\ and \eemumug, larger than 3\%. \pho\ generates around 25\% more additional ISR photons emitted close to one of the beams compared to what is observed in data. By contrast, the \afk\ generator gives a mostly accurate description of data at NLO and NNLO, with a slightly high data/MC ratio at NNLO.

The consequences for past $\pi^+\pi^-(\gamma)$ cross section measurements vary depending on the experiment. The previous \babar\ result~\cite{babar1,babar2} is unaffected as NLO and higher orders were already included in the analysis. The event acceptance, determined with \pho, must be corrected by a factor $(0.3 \pm 0.1) \times 10^{-3}$, negligible compared to the total 0.5\% systematic uncertainty.

Other experiments like BESIII~\cite{bes} and KLOE~\cite{kloe}, however, rely on \pho\ for additional radiations and apply more stringent LO selections. Unaccounted shortcomings from this generator could potentially affect their results and call for larger systematic uncertainties. Hence, the new \babar\ analysis, independent from and potentially more precise than the 2009 study, will be crucial to better understand the tensions between experiments.

\section*{Acknowledgments}

We are grateful for the extraordinary contributions of our PEP-II colleagues in achieving the excellent luminosity and machine conditions that have made this work possible. The success of this project also relies critically on the expertise and dedication of the computing organizations that support \babar, including GridKa, UVic HEP-RC, CC-IN2P3, and CERN. The collaborating institutions wish to thank SLAC for its support and the kind hospitality extended to them. We also wish to acknowledge the important contributions of J.~Dorfan and our deceased colleagues E.~Gabathuler, W.~Innes, D.W.G.S.~Leith, A.~Onuchin, G.~Piredda, and R. F.~Schwitters. This work benefited from funding by the French National Research Agency under contract ANR-22-CE31-0011 and from Laboratoire de physique nucléaire et de hautes énergies (UMR 7585, CNRS/IN2P3, Sorbonne Université, Université Paris-Cité).

\end{document}